\documentclass[usenatbib,useAMS]{mnras}

\usepackage[T1]{fontenc}
\usepackage{ae,aecompl}

\usepackage{graphicx}
\usepackage{amsmath}
\usepackage{amssymb}

\renewcommand{\vec}[1]{\bmath{#1}}
\newcommand{\transpose}[1]{{#1}^{\rmn{T}}}
\newcommand{\mat}[1]{\bmath{\mathsf{#1}}}

\newcommand{\matc}[2]{\mat{#1}^{\rmn{col}}_{#2}}
\newcommand{\matt}[1]{\transpose{\mat{#1}}}

\newcommand{\mati}{\mat{I}}

\newcommand{\iprod}{\cdot}
\newcommand{\oprod}{\otimes}
\newcommand{\eprod}{\circ}
\newcommand{\bigo}[1]{\mathcal{O}\left(#1\right)}

\newcommand{\norm}[1]{\left\|#1\right\|}
\newcommand{\normtwo}[1]{\left\|#1\right\|^2}
\newcommand{\abs}[1]{\left|#1\right|}
\newcommand{\diag}[1]{\rmn{diag}\left(#1\right)}

\newcommand\ionl[3]{\ion{#1}{#2}$\;\lambda${#3}\relax}

\newcommand{\foursym}{\mathcal{F}}
\newcommand{\four}[1]{\foursym{#1}}
\newcommand{\fourp}[1]{\four{\left[#1\right]}}
\newcommand{\ifoursym}{\foursym^{-}}

\newcommand{\ifourp}[1]{\ifoursym{\left[#1\right]}}
\newcommand{\conj}[1]{#1^{*}}

\newcommand{\corr}{\star}
\newcommand{\ltxt}[1]{{\lambda_{\rmn{#1}}}}
\newcommand{\Ntxt}[1]{N_{\mathrm{#1}}}
\newcommand{\lrest}{\ltxt{rest}}
\newcommand{\lobs}{\ltxt{obs}}
\newcommand{\ncomp}{\Ntxt{T}}
\newcommand{\sgn}[1]{\mathrm{sgn}\left(#1\right)}
\newcommand{\erf}[1]{\mathrm{erf}\left(#1\right)}

\newcommand{\shifted}[1]{\tilde{#1}}

\begin{document}

\title[Linearithmic weighted phase correlation]{Redshift determination through weighted phase correlation: a linearithmic implementation}

\author[L. Delchambre]{L. Delchambre$^{1}$\thanks{E-mail:	ldelchambre@ulg.ac.be} \\
$^{1}$ Institut d'Astrophysique et de G\'eophysique, Universit\'e de Li\`ege, All\'ee du 6 Ao\^ut 17, B-4000 Sart Tilman (Li\`ege), Belgique}

\date{Accepted ???. Received ???; in original form ???}

\pagerange{\pageref{firstpage}--\pageref{lastpage}} \pubyear{2016}

\maketitle

\label{firstpage}

\begin{abstract}
We present a new algorithm having a time complexity of $\bigo{N \log N}$ and designed to retrieve the phase at which an input signal and a set of not necessarily orthogonal templates match at best in a weighted chi-squared sense. The proposed implementation is based on an orthogonalization algorithm and thus also benefits from a high numerical stability.  We successfully apply this method to the redshift determination of quasars from the twelfth Sloan Digital Sky Survey (SDSS) quasar catalog and derive the proper spectral reduction and redshift selection methods. Also provided are the derivations of the redshift uncertainty and of the associated confidence. Results of this application are comparable to the performances of the SDSS pipeline while not having a quadratic time dependency.
\end{abstract}


\section{Introduction}
\label{sec:intro}

	The advent of extremely large spectroscopic surveys like the Sloan Digital Sky Survey (SDSS) that includes more than $2\times10^6$ high resolution spectra over 5200 deg$^2$ of the sky \citep{alam2015} or the Gaia space mission that will provide, by the end of 2018, $150\times10^6$ low resolution spectra \citep{debruijne2012} provide us with unique opportunities to have a statistical view on the kind of objects present in the universe along with some of their fundamental characteristics. These play a key role in the answer to some of the currently most important astrophysical questions like the evolution scenarios of the galaxy; of the universe or its accelerated expansion \citep{aubourg2014,perryman2001}.
	
	Along with these large surveys comes an impressive continuous flow of data that has to be treated right on time through huge dedicated processing centers. One of the most important tasks amongst the spectral reduction processes stands in the objects classification and in their astrophysical parameters (APs) determination. More specifically, in the case of extragalactic objects, these informations critically depend on the availability of reliable redshift estimates.
	
	Redshift determination, even if apparently straightforward, is in practice a challenging problem for which numerous solutions have been proposed:
\begin{enumerate}
\item{Visual inspection procedures:} a skilled observer can efficiently guess the APs of any object and can deal with any unexpected cases like corrupted/missing emission lines; spectra superposition or non-physical solutions. Obviously, this choice is unbearable for large surveys though the analysis of any sufficiently large subset is invaluable as it can serve as input to sophisticated computer algorithms that will try to mimic this human expertise. This is the solution undertaken by \cite{paris2015} regarding the redshift of quasars and accordingly, it will be used along this paper as the default quasar spectral library.
\item Matching of spectral lines: this method consists in extracting some significant patterns out of the input spectra and then trying to match them to known emission/absorption lines. This procedure has been used for a long time but has been shown to be restricted to relatively high signal-to-noise ratio spectra \citep{machado2013}.
\item Computer learning methods: the goal is here to make the algorithm guess the relations that exist between some characteristics of already-reduced objects (e.g. observed wavelengths and fluxes), and the parameters of interest (e.g. redshift coming from a visual inspection procedure). The aim being then to apply these relations to the case of objects whose parameters are still unknown. Interested readers may find in \cite{bishop2006} the descriptions of many such algorithms.
Note that, depending on its complexity, the guessed relation may be non-physical and hard to interpret leading to suboptimal or potentially unrealistic predictions. This is the reason why these should preferably be used for the case of highly non-linear problems for which no other--fast--solution exists.
\item Phase correlation: the idea is here to find the optimal correlation of a given observation against one or more templates in order to determine its redshift. Based upon the ability of these templates to match the observations; from the physical nature of this solution and from the shortcomings of the previously mentioned alternatives, we will consider it to be the most trustworthy automated procedure for redshift determination.
\end{enumerate}

	Based on the work of \cite{brault1971}, \cite{simkin1974} first suggested the use of the Fast Fourier Transform (FFT) as an efficient way of finding the redshift of galaxies based on their cross-correlation with a single template. \cite{tonry1979} later derived the formulation  associated with the resulting redshift uncertainties, that were further refined by \cite{heavens1993}. Finally, \cite{glazebrook1997} generalized the cross-correlation technique to the case of templates coming from the principal components analysis (PCA) decomposition of spectral libraries. Although being currently the most widespread technique for redshift determination, the latter actually suffers from some well-known drawbacks (see section \ref{sec:pcorr_pract}). The solution to these problems comes from the use of a weighting scheme associated with the observed spectrum as implemented in \cite{bolton2012}. Unfortunately, this solution has a quadratic time dependency that makes it fairly time consuming.

	The method proposed in the present work overcomes this high numerical complexity and was developed in the framework of the Gaia astrophysical parameters inference system \citep{cbj2013} and more specifically within the field of the quasar classification module  (QSOC) whose goal is to find the APs associated with the quasars that Gaia will detect. In this domain, the time constraints imposed by the Gaia mission restricted us to the use of computer learning methods but in the end, the advent of this new method will allow us to predict fair and fast redshift estimates for the upcoming Gaia data releases.
	
	Section \ref{sec:notation} explains the conventions used along this paper. Section \ref{sec:pcorr_pca} makes a brief review of the phase correlation and PCA techniques aimed at better understanding their main limitations. We have developed a fast solution to the problem of the weighted phase correlation in Section \ref{sec:wcorr}. Tests against real cases are then performed within Section \ref{sec:application} while extensions of the presented algorithm are discussed in Section \ref{sec:discussion}. Finally, we conclude in Section \ref{sec:conclusion}.

\section{Notation}
\label{sec:notation}

This paper uses the following notations: vectors are in bold italic, $\vec{x}$; $x_i$ being the element $i$ of the vector $\vec{x}$. Matrices are in uppercase boldface or are explicitly stated; i.e. $\mat{X}$ from which the $i$th column will be denoted by $\matc{X}{i}$ and the element at row $i$, column $j$ will be denoted by $\mat{X}_{ij}$. In the following, we will consider the problem of finding the optimal offset between an observed spectrum composed of $N_s$ samples and $\ncomp$ templates of size $N_p$ by probing various shift estimate, $Z$. By considering the zero-padding necessary in order for these to be properly used within the Fourier domain, we will have that the template matrices, $\mat{P}$ and $\mat{T}$, will be of size $\left(N\times \ncomp\right)$ with $N = N_s + N_p$. Similarly, we will have that the observation vector, $\vec{s}$, will be of size $N$ as well. Note that in order for the redshift to turn into a simple offset, we will have to consider a logarithmic wavelength scale. If not stated otherwise, matrices and vectors having a tilde on top of them (e.g. $\mat{\shifted{T}}$) will be specific to a given shift try, $Z \in 0\cdots N-1$. Amongst commonly used operators, $\vec{a} \iprod \vec{b}$ denotes the inner-product of $\vec{a}$ and $\vec{b}$; $\vec{a} \oprod \vec{b}$, their outer-product; $\norm{\vec{a}}$, the Euclidian norm of $\vec{a}$ and $\conj{\vec{a}}$ its complex conjugate. Finally, $\fourp{\vec{x}}$ and $\ifourp{\vec{x}}$ respectively corresponds to the discrete Fourier transform (hereafter DFT) and inverse DFT of $\vec{x}$.

\section{Phase correlation using PCA}
\label{sec:pcorr_pca}

	As already stated, the most commonly used technique for QSO redshift determination consists in finding the best correlation of an observed spectrum against templates coming from the PCA decomposition of a restframe spectral library. More specifically, these are based on spectra sampled on a uniform logarithmic wavelength scale such that the observed wavelength, $\lobs$, can be related to the restframe wavelength, $\lrest$, through the QSO redshift, $z$, as a simple offset:
\begin{equation}
\label{eq:log_lobs_lrest}
\log \lobs = \log \lrest + \log (z+1).
\end{equation}

	In the following, we make a brief review of the two above-mentioned techniques that should provide the reader insights about their way of working and aimed at better understanding their main limitations regarding the redshift estimation of QSOs.

\subsection{Principal components analysis}
\label{sec:pca}

	PCA is a well-known technique designed to extract a set of templates --the principal components-- from a typically huge set of data while keeping most of its variance \citep{pearson1901}. These principal components will then be those that are the best suited in order to highlight the most important patterns out of the input data set.
	
	Mathematically, the goal of the PCA is to find a decomposition of an input matrix $\mat{X}$, from which we have subtracted the mean observation, into
\begin{equation}
\label{eq:pca_decomp}
\mat{X} = \mat{P}\mat{C},
\end{equation}
such that
\begin{equation}
\label{eq:pca_eigen}
\mat{D} = \matt{P} \mat{X} \transpose{\mat{X}} \mat{P} = \matt{P} \mat{\sigma^2} \mat{P}
\end{equation}
is diagonal and for which
\begin{equation}
\label{eq:pca_eigen_order}
\mat{D}_i \geq \mat{D}_j; \; \forall i < j.
\end{equation}
$\mat{P}$, the matrix of the eigenvectors of $\mat{\sigma^2}$, is then called the matrix of principal components; $\mat{C}$ is the associated matrix of principal coefficients and $\mat{D}_i$'s are the eigenvalues of the covariance matrix, $\mat{\sigma^2}$. Note that according to the spectral theorem\footnote{Any real symmetric matrix is diagonalized by a matrix of its eigenvectors.}, $\mat{P}$ will be orthonormal such that we have
\begin{equation}
\label{eq:pca_coeffs}
\mat{C} = \matt{P} \mat{X}.
\end{equation}
From this orthonormality and from equation \ref{eq:pca_eigen_order}, we will have that the linear combination of the first principal components of $\mat{P}$ with the associated principal coefficients of $\mat{C}$ will constitute the best linear combination in order to fit $\mat{X}$ in a least squares sense.
	
	An illustrative example of PCA decomposition is given in figure \ref{fig:pca_example}. The latter is based on spectra covering the restframe wavelength range 1100--2000\r{A} coming from the SDSS DR12 quasar catalog \citep{paris2015}. Notice how the main QSOs emission lines are modelled by the various components as a way to grab the variance coming from the great diversity of shapes encountered within the spectral library. Readers willing more information on the PCA decomposition are invited to read \cite{jolliffe2002} for a deep analysis of the technique or \cite{schlens2014} for an accessible tutorial.
	
	The application of this technique to the analysis of QSO spectra was first covered by \cite{francis1992}; \cite{yip2004} later adapted it to the case of the SDSS DR1 quasars classification and redshift determination while \cite{cabanac2002} did a similar work based upon spectra coming from the Large Zenith Telescope survey whose spectral resolution ($\lambda/\Delta\lambda \sim 40$) is in the same order of magnitude as the one of the red and blue photometers of Gaia \citep{cbj2013}.
	
\begin{figure}
\includegraphics[width=0.5\textwidth]{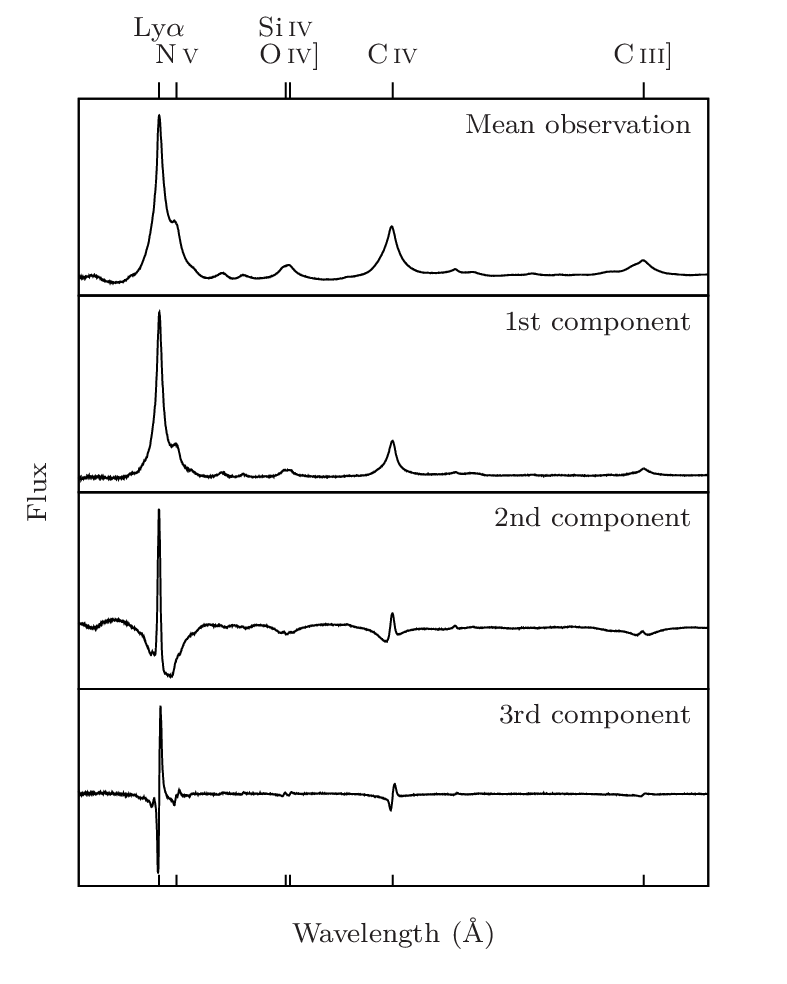}
\caption{Restframe mean observation and first principal components coming from the PCA decomposition of TypeI/II SDSS DR12 quasars spectra having $2.3 \leq z \leq 2.4$ (24939 spectra).}
\label{fig:pca_example}
\end{figure}

\subsubsection{Weighted PCA}
\label{sec:wpca}

	One of the main limitations of the classical PCA method is that it does not make any distinction between variance coming from noise and variance coming from a genuine signal. Furthermore, in its naive form, it does not know how to deal with missing data. This last point is particularly crucial in the field of high-redshift surveys where the observed wavelength ranges may not overlap from object to object. 

	A straightforward approach so as to avoid these shortcomings stands in the use of a weighting scheme that allows each flux within each spectrum to come along with its own uncertainty while performing the PCA decomposition. Such a fully-weighted PCA (WPCA) method was first described in the astronomical literature by \cite{tsalmantza2012} and was later refined by \cite{bailey2012}. In the field of the present study, we will use the implementation described in \cite{delchambre2015}, this choice mainly comes from its high numerical stability. This method is based on the diagonalization of the weighted variance-covariance matrix as defined by
\begin{equation}
\label{eq:wpca_covmat}
\mat{\sigma^2} = \frac{\left(\mat{X}\eprod\mat{W}\right)
\transpose{\left(\mat{X}\eprod\mat{W}\right)}}{\mat{W}\matt{W}},
\end{equation}
where $\eprod$ represents the element-wise product of two matrices and where $\mat{X}$ is supposed to have a weighted mean observation of zero. The decomposition of $\mat{\sigma^2}$ into a diagonal matrix of eigenvalues, $\mat{D}$, and a matrix of orthonormal principal components, $\mat{P}$, being then performed using either a combination of two spectral decomposition methods, namely the power iteration method followed by the Rayleigh quotient iteration one, or by the use of the singular value decomposition (SVD). This technique allows us to retrieve the fairest components (i.e. those for which uncertainties are taken into account) without having to worry about missing data: this case being the limiting case of weights equal to zero. Consequently, this method will be used through the rest of this document as the default process in order to retrieve the principal components.

\subsection{Phase correlation}
\label{sec:pcorr}

	The goal of the phase correlation algorithm is to find the optimal shift between a set of orthonormal templates --or a sole unit-length template--, $\mat{P}$, and a given observation, $\vec{s}$, that has been shifted relatively to $\mat{P}$. The way to proceed is to compute for each potential shift, $Z$, the linear least-squares solution of the shifted templates, $\mat{\shifted{P}}_{ij} \equiv \mat{P}_{(i+Z)j}$, against the observation such as to find the offset having the minimal resulting chi-square. More concisely, this is equivalent to find the minimal shift-dependent chi-square as defined by
\begin{equation}
\label{eq:pcorr_chi2}
\chi^2(Z) = \normtwo{\vec{s} - \mat{\shifted{P}} \vec{a}(Z)},
\end{equation}
where $\vec{a}(Z)$ contains the optimal linear coefficients in order to fit $\vec{s}$ based on $\mat{\shifted{P}}$.

	Extending the work of \cite{simkin1974}, \cite{glazebrook1997} noticed that in the case of orthonormal templates, like the PCA principal components, equation \ref{eq:pcorr_chi2} becomes
\begin{equation}
\label{eq:pcorr_chi2_orth}
\chi^2(Z) = \normtwo{\vec{s}} - \normtwo{\vec{a}(Z)}.
\end{equation}
Consequently, equation $\ref{eq:pcorr_chi2}$ will be minimal for an associated maximal $\normtwo{\vec{a}(Z)}$. Moreover, due to the orthonormality of $\mat{P}$, we will have that
\begin{equation}
\label{eq:pcorr_az_p}
\vec{a}(Z) = \matt{\shifted{P}} \vec{s}.
\end{equation}
More specifically, regarding the $i$th linear coefficient, $a_i(Z)$, we will have that
\begin{equation}
\label{eq:pcorr_aiz_p}
a_i(Z) = \sum_j \mat{P}_{(j+Z)i} s_j = \left(\matc{P}{i} \corr \vec{s}\right)_Z.
\end{equation}
We recognize equation \ref{eq:pcorr_aiz_p} as being the correlation of the vector $\matc{P}{i}$ with $\vec{s}$ that can hence be efficiently computed in the Fourier domain. Interested readers may find in \cite{brault1971} exhaustive hints about the practicalities surrounding the Fourier implementation of equation \ref{eq:pcorr_aiz_p}. Let us just point out that both vectors, $\matc{P}{i}$ and $\vec{s}$ have to be extended and zero-padded such as to deal with the periodic nature of the DFT. Note that in the rest of this document, the curve obtained after evaluating $\normtwo{\vec{a}(Z)}$ at each $Z$ will be termed the \textit{cross-correlation function} (CCF).

	A sub-sampling precision on the offset can be gained by considering the fit of a continuous function in the vicinity of the maximal peak of the discrete CCF. \cite{simkin1974} supposed this peak to be Gaussian profiled, but in the aim of having a model-independent estimate of $Z$, we will follow \cite{tonry1979} and use a quadratic curve fitting that will allow us to take into account potential asymmetries in the fitted peak.

\subsubsection{Practicalities}
\label{sec:pcorr_pract}

	Some issues highlighted in \cite{glazebrook1997} are the subtraction of the QSO continuum and of the restframe mean spectrum from the observed spectrum. The first issue was here solved by the use of a dedicated method that allows us to fit the QSO continuum in a fast and redshift-independent way. This method will be further described in section \ref{sec:app_proc_desc}. The second issue is often overcome by omitting the subtraction of the mean spectrum from the input dataset. We have to note that this omission typically degrades the ability of the PCA decomposition to extract the most significant patterns out of this input dataset. Another solution would have been to alter the mean spectrum such as to make it orthonormal to the template components, $\mat{P}$, --thanks to the use of a Gram-Schmidt orthogonalization process \citep{press2002} for example-- and to further consider it as being an additional template. This solution will be adopted here for the use of the phase correlation algorithm.

\begin{figure*}
\includegraphics{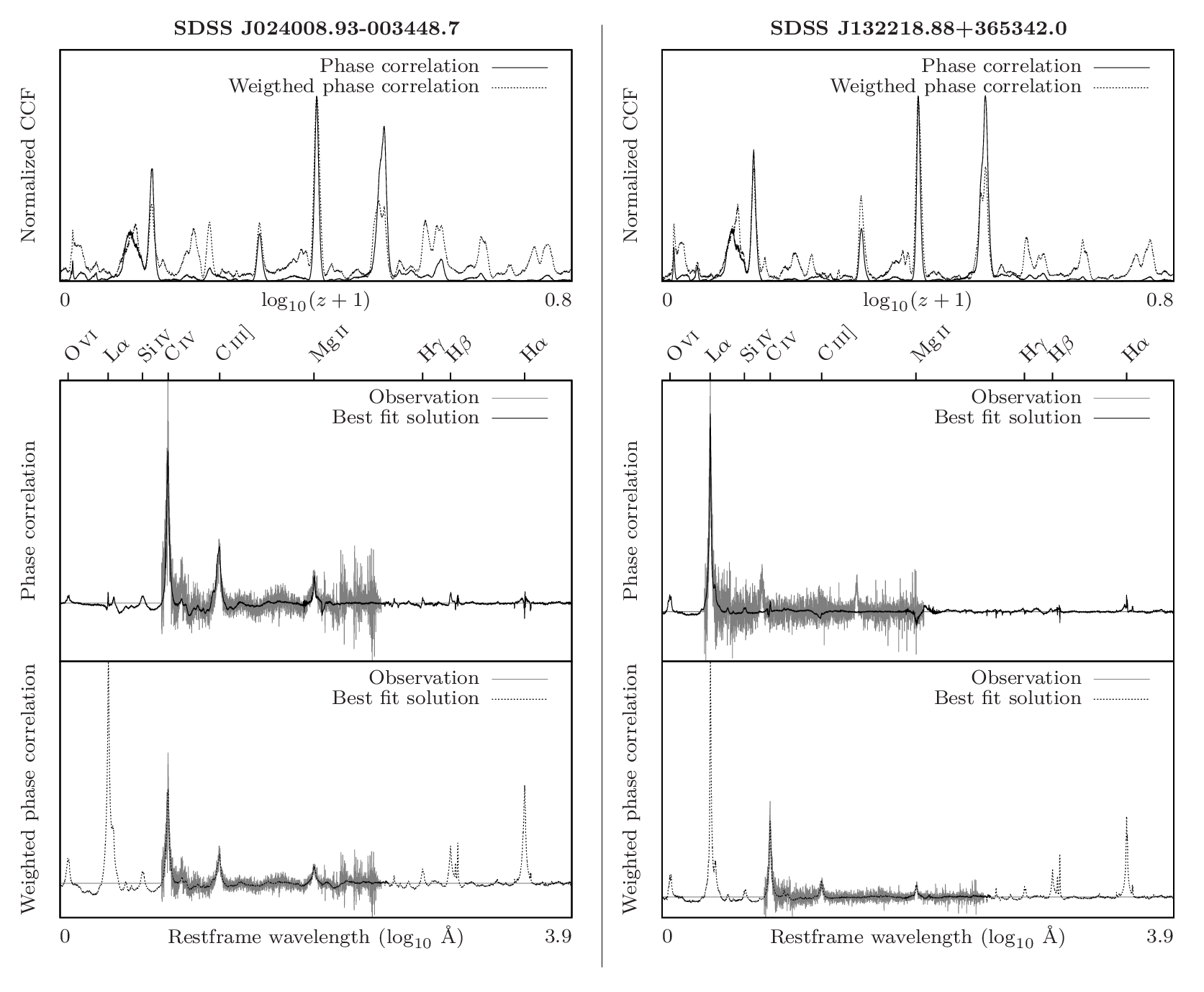}
\caption{Examples of cross-correlation functions coming from the phase correlation and weighted phase correlation of two quasar spectra against the ten first SDSS DR12 principal components plus mean observation (top). The best fit solutions associated with the maximal peak of the phase correlation (middle) and of the weighted phase correlation (bottom) are also provided.}
\label{fig:corr_comparison}
\end{figure*}	
	
	Finally, the major drawback of the implementation of \cite{glazebrook1997} stands in the fact that the observed spectra typically span only a small part of the template spectra such that the CCF will be computed over a substantial number of unknown points. As a consequence, the fit of the input spectra will be disrupted by the `flattening' of the principal components over the unobserved wavelengths. Figure \ref{fig:corr_comparison} illustrates the result of the phase correlation algorithm along with the best-fit solution associated with the maximal peak of the CCF. Notice how the solutions are flattened over unobserved wavelengths. More precisely, considering the observation of "SDSS J024008.93-003448.7", the Ly$\alpha$, H$\alpha$ and H$\beta$ emission lines are strongly damped despite the fact that the optimal shift was found while for the observation of "SDSS J132218.88+365342.0", this `flattening' has led to an ambiguity in the CCF that leads to an erroneous shift estimate. Additionally, uncertainties about the observed fluxes are often available and will not be used within this implementation.

\section{Weighted phase correlation}
\label{sec:wcorr}

	With the aim of dealing efficiently with the previously mentioned problem of unobserved wavelengths and of neglected uncertainties, we will use a $\chi^2$ formulation similar to equation \ref{eq:pcorr_chi2}, but whose fluxes are weighted according to the observed spectrum wavelengths. Also, we will drop the orthonormality constraint on the fitted templates since, in anyway, the previously mentioned weighting will break it down. We will then have the following objective formula:
\begin{equation}
\chi^2(Z) = \left\|\mat{W}\vec{s}-\mat{W}\mat{\shifted{T}}\vec{a}(Z)\right\|^2 = \left\|\vec{y}-\mat{\shifted{X}}\vec{a}(Z)\right\|^2,
\label{eq:wcorr_chi2}
\end{equation}
where $\mat{W}$ is the diagonal matrix of weights associated with $\vec{s}$ and $\mat{\shifted{T}}_{ij} \equiv \mat{T}_{(i+Z)j}$ is the shifted matrix of --not necessarily orthonormal-- template observations. The fastest solution in order to minimize equation \ref{eq:wcorr_chi2} for a given $Z$ stands in the use of a Cholesky decomposition of the design matrix, $\matt{\shifted{X}}\mat{\shifted{X}}$, followed by a forward-backward substitution associated with the image vector $\matt{\shifted{X}}\vec{y}$ \citep{press2002}. We have to note that this approach is known to suffer from numerical instabilities \citep{golub1996,press2002} and is solely provided here as a comparison point regarding its computational performances. Practically, slower but more stable methods based on the orthogonalization of $\mat{\shifted{X}}$ should be preferred.

	In a computational point of view, the evaluation of equation \ref{eq:wcorr_chi2} for each $Z$ will require $\bigo{N^2}$ flops\footnote{Floating operations}\footnote{The interested reader may find in \cite{golub1996} informations and references about the various algorithmic complexities used along this document.}, the latter being mainly dedicated to the building of the design matrices. This relatively high complexity constitutes the main limitation of this implementation and makes it unaffordable for the tight processing of a large survey like Gaia. Nonetheless, it has proven to provide fair redshift estimates and is currently being effectively used in the SDSS-III spectral classification redshift measurement pipeline with a singular value decomposition (SVD) of $\mat{\shifted{X}}$ advantageously replacing the Cholesky decomposition of the design matrix \citep{bolton2012}.

\subsection{Orthogonal decomposition approach}
\label{sec:wcorr_orth}

	The previous section points out the risks encountered while using a naive approach for solving the normal equations associated with equation \ref{eq:wcorr_chi2}. In this optics, let us explore the effect of the orthogonalization of $\mat{\shifted{X}}$ on the latter equation. For this purpose, let us detail the QR decomposition of $\mat{\shifted{X}} = \mat{Q} \mat{R}$\footnote{Note that we dropped the upper tilde for clarity purpose}, that is such that
\begin{equation}
\matt{Q} \mat{\shifted{X}} = \mat{Q_{\ncomp-1}} \cdots \mat{Q_1} \mat{\shifted{X}} = \mat{Q_{\ncomp-1}} \cdots \mat{Q_i} \mat{\shifted{X}_i} = \mat{R} ,
\label{eq:wcorr_qr}
\end{equation}
where $\mat{R}$ is an upper triangular matrix of size $(N \times \ncomp)$ and where each $\mat{Q_i}$ is an Householder reflection matrix designed to annihilate the elements below the $i$th row of the $i$th column of $\mat{\shifted{X}_i}$ \citep{press2002}. More precisely, given $\mat{\shifted{X}_i^\prime}$, the not-already upper-triangular part of $\mat{\shifted{X}_i}$,
we will have
\begin{equation}
\mat{Q}_i = \left(
\begin{array}{cc}
\mati & \mat{0} \\
\mat{0} & \mati - 2 \vec{v_i} \oprod \vec{v_i} / \left\|\vec{v_i}\right\|^2
\end{array}
\right) =\left(
\begin{array}{cc}
\mati & \mat{0} \\
\mat{0} & \mat{Q_i^\prime}
\end{array}
\right)
\label{eq:wcorr_qi}
\end{equation}
with 
\begin{equation}
\vec{v_i} = \vec{x_i} \pm \norm{\vec{x_i}} \vec{e_1};
\label{eq:wcorr_vi}
\end{equation}
$\vec{x_i}$ being the first column of $\mat{\shifted{X}_i^\prime}$ and $\vec{e_1}$ being the first row of the identity matrix. For numerical stability reasons, the choice between subtraction and addition in equation \ref{eq:wcorr_vi} should be matched to the sign of the first element of $\vec{x_i}$ \citep{press2002}. 

	By using such a decomposition, we will have that equation \ref{eq:wcorr_chi2} becomes
\begin{equation}
\chi^2(Z) = \left\|\vec{y}-\mat{Q}\mat{R}\vec{a}(Z)\right\|^2 = \left\|\vec{y}-\mat{Q}\vec{b}(Z)\right\|^2,
\label{eq:wcorr_chi2_orth}
\end{equation}
with the last $N-\ncomp$ elements of $\vec{b}(Z)$ being zeros. The point is now to recognize equation \ref{eq:wcorr_chi2_orth} as being the weighted counterpart of equation \ref{eq:pcorr_chi2} such that the first $\ncomp$ elements of $\vec{b}(Z)$ will be equal to the first $\ncomp$ elements of $\matt{Q} \vec{y}$, whose computation can be efficiently performed by successive multiplication of each of the $\mat{Q_i}$  with the associated $
\vec{y_i} \equiv \mat{Q_{i-1}} \cdots \mat{Q_1} \vec{y} = \left(b_1(Z) \cdots b_{i-1}(Z) \vec{y_i^\prime}\right)$,
rather than by explicitly computing the general $\mat{Q}$ matrix. This efficiency mainly comes from the fact that:

\begin{enumerate}
\item We do not need to explicitly compute any $\mat{Q_i^\prime}$, since we will have that the $j$th column of the product $\mat{Q_i^\prime} \mat{\shifted{X}_i^\prime}$ will be given by
\begin{equation}
\left(\mat{Q_i^\prime}\mat{\shifted{X}_i^\prime}\right)_j^{\mathrm{col}} = \left(\mat{\shifted{X}_i^\prime}\right)_j^{\mathrm{col}} - 2 \dfrac{\vec{v_i} \iprod \left(\mat{\shifted{X}_i^\prime}\right)_j^{\mathrm{col}}}{\vec{v_i} \iprod \vec{v_i}} \vec{v_i},
\label{eq:wcorr_qip_xip}
\end{equation} 
and similarly,
\begin{equation}
\mat{Q_i^\prime}\vec{y_i^\prime} = \vec{y_i^\prime} - 2 \dfrac{\vec{v_i} \iprod \vec{y_i^\prime}}{\vec{v_i} \iprod \vec{v_i}} \vec{v_i}.
\label{eq:wcorr_qip_yip}
\end{equation}
That is: the computation of $\mat{Q_i^\prime}\vec{y_i^\prime}$ and of any column of the products $\mat{Q_i^\prime} \mat{\shifted{X}_i^\prime}$ is now reduced to a single inner product (the product $\vec{v_i} \iprod \vec{v_i}$ being common to all multiplications, it can be pre-computed) and to a single vector subtraction.
\item We do not need to compute any $\mat{R}_{ij}$. Differently stated, we do not need to compute the first row nor the first column of any $\mat{Q_i^\prime}\mat{\shifted{X}_i^\prime}$.
\end{enumerate}

	This implementation, termed `factorized QR algorithm', has a total complexity which can compete with the Cholesky solution of the normal equations while gaining in numerical stability. But practically it is of low interest for us since it remains a quadratic problem that is consequently out of the time processing required by the Gaia tight data reduction.
	
	Let us note that the equation \ref{eq:wcorr_chi2_orth} still provides us with a weighted formulation of the CCF, that is $\left\|\vec{b}(Z)\right\|^2$, such that we can already investigate the effects of the weighting on the best fit solutions at its maximal peak and on the CCF itself. As illustrated in figure \ref{fig:corr_comparison}, the fitted spectra do no longer exhibit border flattening and thanks to this, the maximal peaks are now clearly identified. More particularly, regarding the observation of "SDSS J132218.88+365342.0", the optimal peak of the CCF turns out to be unambiguously identified thanks to the use of this weighted formulation of the phase correlation.

\subsubsection{Factorized QR algorithm with lookup tables}
\label{sec:wcorr_orth_factorized_lookup}

	The quadratic nature of the factorized QR algorithm comes from the large amount of inner products involved in the computation of the first $\ncomp$ elements of each $\vec{b}(Z)$. More specifically, by developing each inner product coming from equations \ref{eq:wcorr_qip_xip} and \ref{eq:wcorr_qip_yip} in the case of the initial reduction $\mat{\shifted{X}_2} \equiv \mat{Q_1}\mat{\shifted{X}}$ and associated image production $\vec{y_2} \equiv \mat{Q_1}\vec{y}$, we get
\begin{eqnarray}
\vec{v_1} \iprod \vec{v_1} & = & 2 \alpha \left(\alpha + \mat{\shifted{X}}_{11}\right),
\label{eq:wcorr_v1_v1} \\
\vec{v_1} \iprod \vec{y} & = & \alpha y_1 + \matc{\shifted{X}}{1} \iprod \vec{y}
\label{eq:wcorr_v1_y},\\
\vec{v_1} \iprod \matc{\shifted{X}}{j} & = & \alpha \mat{\shifted{X}}_{1j} + \matc{\shifted{X}}{1} \iprod \matc{\shifted{X}}{j}
\label{eq:wcorr_v1_Xj}
\end{eqnarray}
with $\alpha = \sgn{\mat{\shifted{X}}_{11}} \left(\matc{\shifted{X}}{1} \iprod \matc{\shifted{X}}{1}\right)^{\frac{1}{2}}$. At this point, it should be noted that 
\begin{equation}
\matc{\shifted{X}}{i} \iprod \matc{\shifted{X}}{j} = \vec{w}^2 \iprod \left(\matc{\shifted{T}}{i} \eprod \matc{\shifted{T}}{j}\right) = \sum_{k=1}^N w_k^2 \left(\matc{T}{i} \eprod \matc{T}{j}\right)_{k+Z}
\label{eq:wcorr_Xi_Xj}
\end{equation}
and that
\begin{equation}
\matc{\shifted{X}}{i} \iprod \vec{y} = \left(\vec{w}^2 \eprod \vec{s}\right) \iprod \matc{\shifted{T}}{i} = \sum_{k=1}^N w_k^2 s_k \mat{T}_{(k+Z)i},
\label{eq:wcorr_Xi_y}
\end{equation}
with $\vec{w} \equiv \diag{\mat{W}}$. We can readily see that equations \ref{eq:wcorr_Xi_Xj} and \ref{eq:wcorr_Xi_y} can be efficiently computed in the Fourier domain. In order to take benefits from it, let us define the \textit{lookup table} of the inner products of $\mat{\shifted{X}}$ with itself as
\begin{equation}
\mat{\shifted{L}}_{ij} = \matc{\shifted{X}}{i} \iprod \matc{\shifted{X}}{j} = \ifourp{\conj{\fourp{\matc{T}{i}\eprod\matc{T}{j}}} \eprod \fourp{\mat{W}^2}}_Z,
\label{eq:wcorr_lookup}
\end{equation}
and the one containing the inner products of $\mat{\shifted{X}}$ with $\vec{y}$ as
\begin{equation}
\vec{\shifted{l}}_i = \matc{\shifted{X}}{i} \iprod \vec{y} = \ifourp{\conj{\fourp{\matc{T}{i}}} \eprod \fourp{\mat{W}^2\vec{s}}}_Z.
\label{eq:wcorr_ylookup}
\end{equation}
Note that in the latter equations, $\conj{\fourp{\matc{T}{i}\eprod\matc{T}{j}}}$ and $\conj{\fourp{\matc{T}{i}}}$ are template-specific and can hence be computed in advance.

	Explicitly stated, these lookup tables allow us to have for any shift estimates, $Z$, an instantaneous evaluation of all the inner products associated with the initial reduction process. Furthermore, thanks to the $\mat{Q_1}$ orthonormality, we will have that the lookup tables associated with $\mat{\shifted{X_2}}$ will be also given by $\mat{\shifted{L}}$ and $\vec{\shifted{l}}$. Consequently, we can easily compute the inner product of $\mat{\shifted{X}_2^\prime}$ with itself based on $\mat{\shifted{L}}$ as
\begin{equation}
\left(\mat{\shifted{X}_2^\prime}\right)_i^\mathrm{col} \iprod \left(\mat{\shifted{X}_2^\prime}\right)_j^\mathrm{col} = \mat{\shifted{L}}_{ij} - \mat{R}_{1i} \mat{R}_{1j}; \qquad \forall i,j
\label{eq:wcorr_lookup_update}
\end{equation}
and in the same way, we can compute the inner products of $\mat{\shifted{X}_2^\prime}$ with $\vec{y_2^\prime}$ based on $\vec{\shifted{l}}$ as
\begin{equation}
\left(\mat{\shifted{X}_2^\prime}\right)_i^\mathrm{col} \iprod \vec{y_2^\prime} = \vec{\shifted{l}}_i - \mat{R}_{1i} b_1(Z); \qquad \forall i.
\label{eq:wcorr_ylookup_update}
\end{equation}
Equations \ref{eq:wcorr_lookup_update} and \ref{eq:wcorr_ylookup_update}  will allow us to recursively process each subsequent $\mat{\shifted{X}_i^\prime}$ and $\vec{y_i^\prime}$ in a way similar to the one used to produce $\mat{\shifted{X}_2}$ and $\vec{y_2}$ and will be referred to as the \textit{lookup tables update equations}. Finally, let us note that once these lookup tables have been computed, only the first $\ncomp$ rows of $\mat{\shifted{X}}$ and $\vec{y}$ are now needed for the algorithm to run.

	If we suppose now  that $\ncomp \ll N$, then we will have that most of the computation time will be spent in the building of the initial values of the lookup tables (equations \ref{eq:wcorr_lookup} and \ref{eq:wcorr_ylookup}). More precisely these will crudely correspond to the DFT of $\vec{w^2}$ and of $\vec{w^2} \eprod \vec{s}$; their vector multiplication  with each combination of the templates plus the inverse transforms leading to these initial values. Despite the fact that the previous derivation is a bit coarse, it still assesses the linearithmic (i.e. $\bigo{N \log N}$) behaviour of the presented algorithm. Regarding now the specific problem of the QSO redshift determination within the Gaia mission (expected to be $N=10^4$, $\ncomp=10$), tests performed on a 2,4Ghz CPU provide execution times of $180.35 \pm 6.76$ seconds for the normal equations solution compared to $0.173 \pm 0.002$ second for our implementation; these become respectively $4.95 \pm 0.19$ hours compared to $1.81 \pm 0.02$ second for the case of $N=10^5$ and $\ncomp=10$. Finally, let us note that the proposed algorithm can be easily implemented in parallel given the fact that the estimation of each $\chi^2(Z)$ can be separately performed. As a consequence, the execution time can be scaled by an arbitrary factor that is inversely proportional to the number of running processes.

\section{Application}
\label{sec:application}

	Unsurprisingly, the performance of the presented method was assessed on type I/II QSOs coming from the SDSS DR12 quasar catalog \citep{paris2015}. The choice of this catalogue comes from the fact that all spectra contained therein were visually inspected and can hence be considered as being extremely reliable regarding their redshift. Additionally, it is also interesting to note that the latter contains a non-negligible number of 297 301 QSOs that is adequate in order to derive strong statistics.
	
	Due to time constraints and to the need for the WPCA algorithm to have a well covered input space of parameters (i.e. numerous observations), we used a two-fold cross-validation in order to test our method. That is: we split our input catalog into two randomly drawn parts out of which we extract the principal components; then we compute the redshift of spectra belonging to each part based on both the weighted and classical phase correlation algorithms whose inputs are the principal components built on the alternative part. Following is a detailed description of the processes leading to this cross-validation.
	
\subsection{Procedure description}
\label{sec:app_proc_desc}
	
	Raw spectra are generally not readily exploitable. Rather, we have to reduce them such as to get rid of most of the contaminating signals that encompass, for the specific case of this study: deviant points (amongst which night sky emission lines and spectrograph edge effects) and QSOs continuum. Note that since the SDSS DR12 spectra are already sampled on a uniform logarithmic scale, nothing has to be done in order for equation \ref{eq:log_lobs_lrest} to be fulfilled but usually spectra have to be resampled.
	
	The estimation of the QSO continuum turns out to be a challenging problem upon which the quality of the principal components and of the redshift prediction strongly depend \citep{machado2013}. Four broad kinds of approaches have been investigated so far in order to estimate this continuum: (1) the fit of a `damped' power-law function to the observed spectra \citep{ferland1996}; (2) the use of PCA such as to predict the shape of the Ly$\alpha$ forest continuum based on the red part of the spectrum \citep{suzuki2005,paris2011,lee2012}; (3) the modelling of the dependency between the intrinsic QSO continuum and the absorption that it encounters as a mean to extrapolate it \citep{bernardi2003} and (4) through the use of techniques related to the multiresolution analysis \citep{dallaglio2008,machado2013}. We choose to use this last alternative based on the fact that we do not require the resulting continuum to have a physical basis (i.e. the continuum subtraction being rather used as a normalization) and on the fact that we would like to have the most empirical estimation of this continuum. Following \cite{machado2013}, we found that the signature of the continuum clearly stands within the low frequency components of the pyramidal median transform \citep[hereafter PMT]{starck1996} of the input spectrum. In practice, the PMT is computed on a flipped version of the spectrum concatenated with the original version and another flipped version such as to ensure continuity at the border. After taking the inverse transform through a third degree fitting polynomial, we enforce the smoothness of the solution by convolving it with a thousand points-wide Savitzky-Golay filter such as to provide the final continuum. Besides its accuracy, we have to note that the PMT, from its pyramidal nature, has an algorithmic complexity of $\bigo{N\log N}$ and will consequently not degrade the performances of the global process.
	
	After having subtracted the derived continuum, we discard border regions for which either $\lambda < 3800$\r{A} or $\lambda > 9250$\r{A}; we reject $4$\r{A} regions around each significant night sky emission lines and finally we perform a $k$-sigma clipping ($k = 3$, $\sigma = 4$) on the two first scales of the PMT such as to remove extremely deviant points. Finally, we get an estimate of the signal-to-noise ratio (hereafter SNR) of each continuum-subtracted spectrum through the computation of a `noiseless' spectrum coming from the hypothesis that the noise within these spectra is entirely contained within the five first scales of the biorthogonal spline stationary wavelet transform of each spectrum \citep{cohen1992,burrus1997}. Practically, a spline of third degree was used for both analysis and synthesis. Figure \ref{fig:preprocess} illustrates the result of the initial reduction process.
	
\begin{figure*}
\includegraphics{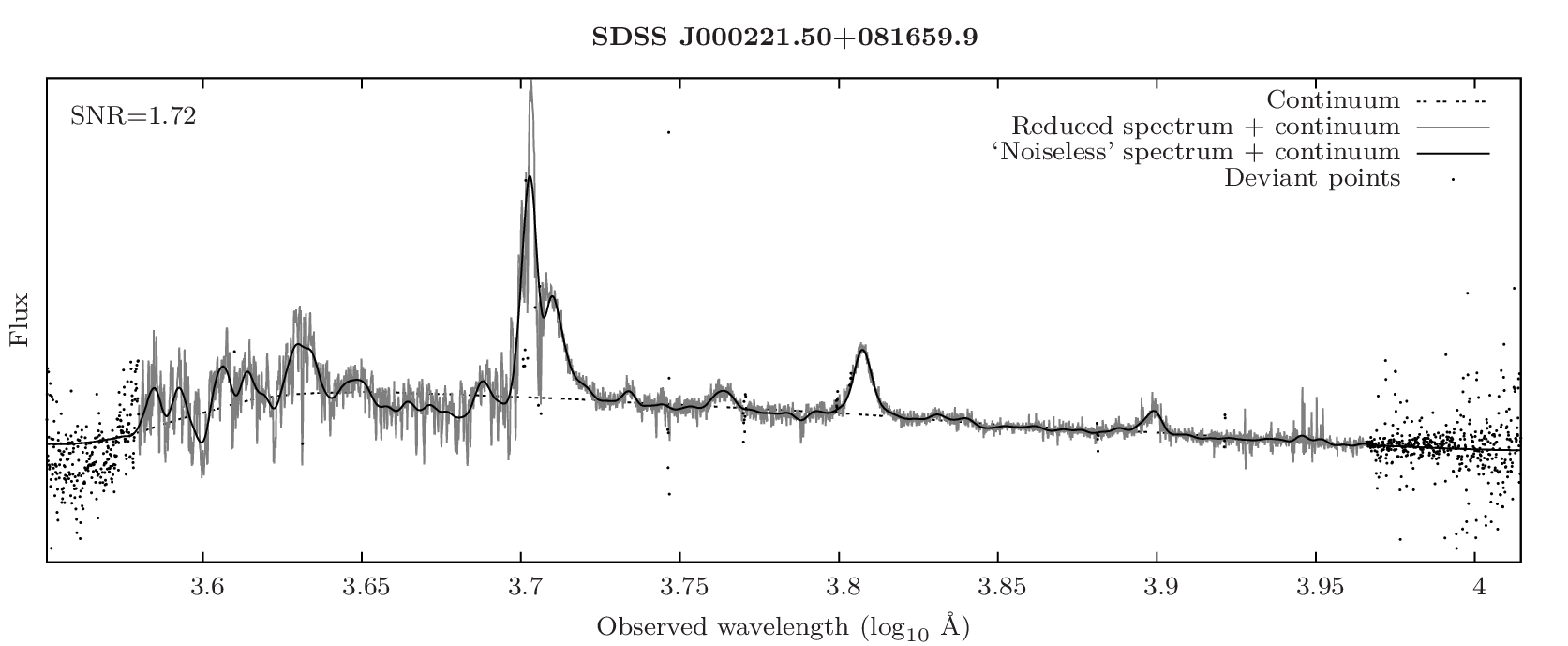}
\caption{Result of the initial reduction process designed to decompose any input spectrum into: a continuum spectrum through the smoothing of the low-frequency components of the PMT; a `noiseless' spectrum through the removal of high-frequency components of the biorthogonal spline stationary wavelet transform and a set of deviants points through a $k$-sigma clipping of the first two scales of the PMT as well as through the removal of night sky emission lines and border regions. Also computed is the SNR of the continuum-subtracted spectrum. The continuum drop occurring at wavelengths shorter than the Ly$\alpha$ limit can be explained by the absorption induced by the intergalactic medium --mainly composed of hydrogen atoms-- located along the line-of-sight towards the QSO under study \citep{petitjean1993}.}
\label{fig:preprocess}
\end{figure*}

	Spectra having an estimated SNR greater than 1 are then set on a common logarithmic wavelength scale with a uniform sampling of $\Delta \log_{10} \lambda = 10^{-4}$, equal to the original sampling of the spectra. The 116 374 resulting spectra are then divided into two equal parts --called learning sets-- each of which is being used to produce the principal components and mean observations associated with each part of the cross-validation process. Resulting from this subdivision, we will have that the input catalogue will be split into two parts --the test sets-- each consisting in 133 860 observations. Note that given the fact that the broad absorption line QSOs are discarded, both sets do not sum up to 293 301 QSOs.

	We then compute the classical and weighted CCF of each spectrum contained within the two test sets based on the mean observation and ten first principal components coming from the alternative learning set. Out of these CCF we extract the five most significant peaks --having a separation of at least $15,000$km s$^{-1}$-- and we fit them with a second order polynomial such as to gain a sub-sampling precision on the predicted peak position. Note that we choose to consider multiple solutions based on the fact that the most significant peaks may not always lead to a physical basis. For example, we might have deep absorption lines either coming from the host galaxy of the quasars or from extragalactic objects being located along the line-of-sight during acquisition and leading to `negative' fitted emission lines. These can definitely prevent the highest peak --the one with the associated minimal $\chi^2$-- from being the effective one. In order to discriminate between these five selected solutions, we define two score measures: $\chi_r^2(z)$, defined as the ratio of the value of the peak associated with the redshift $z$ to the value of the maximal peak and $Z_{\mathrm{score}}(z)$, defined as
\begin{equation}
Z_{\mathrm{score}}(z) = \prod \frac{1}{2} \left[1 + \erf{\frac{e_\lambda}{\sigma(e_\lambda)\sqrt{2}}} \right],
\label{eq:zscore}
\end{equation}
where $e_\lambda$ are the mean values of the emission lines covered by the observed spectrum if we consider it to be at redshift $z$ and where $\sigma(e_\lambda)$ are the associated uncertainties. Note that both $e_\lambda$ and $\sigma(e_\lambda)$ are computed over a range of eleven points surrounding each emission line. We can recognize each term of equation \ref{eq:zscore} as being the cumulative distribution function of a normal distribution of mean zero and variance $\sigma^2(e_\lambda)$ evaluated at $e_\lambda$. The use of equation \ref{eq:zscore} allows us to have a numerical estimate of the ability for a given redshift, $z$, to grab the following chosen QSO emission lines: \ionl{O}{vi}{1033}; Ly$\alpha\;\lambda 1215$; \ionl{N}{v}{1240}; \ionl{Si}{iv}{1396}; \ionl{C}{iv}{1549}; \ionl{C}{iii]}{1908}; \ionl{Mg}{ii}{2797}; H$\gamma\;\lambda 4340$; H$\beta\;\lambda 4861$ and H$\alpha\;\lambda 6562$\r{A}. Typical values of $Z_{\mathrm{score}}(z)$ range from $\sim 1$ for a solution with a clear match of all positive emission lines while it voluntarily penalizes solutions with a match of at least one `negative' emission line by giving them a $Z_{\mathrm{score}}(z) \sim 0$, values in between often occur in low SNR spectra or spectra with strongly damped emission lines. Finally, an error on each estimated peak position is derived and will be further described in section \ref{sec:discussion_zerror}.

	For each spectra, the selection of the optimal redshift out of the five potential ones, $z_1, \cdots, z_5$ for which $1 = \chi_r^2(z_1) \geq \chi_r^2(z_2) \geq \cdots \geq \chi_r^2(z_5)$ and coming either from the classical CCF or from the weighted CCF is done in the following way: if $Z_{\mathrm{score}}(z_1) > 0.8$, then select $z_1$; otherwise choose the shift having the highest $\chi^2_r$ and for which both $Z_{\mathrm{score}}(z_i) > 1-10^{-6}$ and $\chi^2_r(z_i) > 0.8$; otherwise choose the shift having the highest $Z_{\mathrm{score}}$ and for which $\chi_r^2(z_i) > 0.9$. Note that the previous selection and constants therein are purely empirical and based on an iterative visual inspection of misclassified spectrum. This final step provides us with what we thought to be the most probable redshift estimate for a given input spectrum along with the associated uncertainty and a warning flag notifying a failure and/or imprecision in the CCF computation; in the peak identification or in the redshift selection (e.g. all fluxes to zero, low $Z_{\mathrm{score}}$ or less precise uncertainties). 
	
\subsection{Results}
\label{sec:app_results}

\begin{figure}
\includegraphics{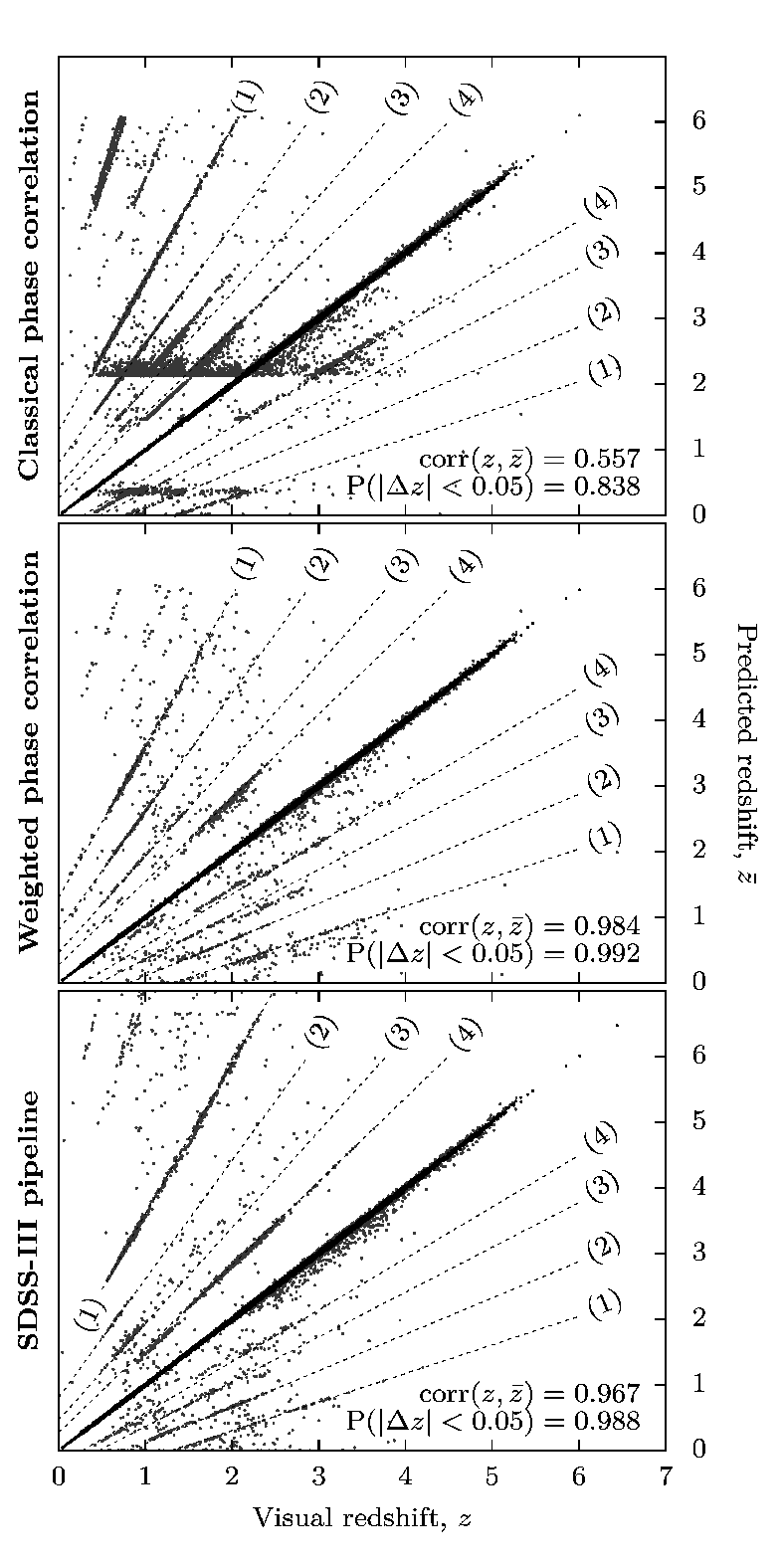}
\caption{Results of the classical phase correlation and weighted phase correlation algorithms based on the cross-validation of observations coming from the SDSS DR12Q quasar catalog plus predictions coming from the SDSS-III pipeline. Provided informations are the correlation factor between $z$ and $\bar{z}$ and $\mathrm{P}(\left|\Delta z\right| < 0.05)$, the ratio of observations having an absolute error lower than $0.05$ (depicted as black dots). Dotted numbered lines correspond to known mismatches between common emission lines: (1) \ion{Mg}{ii} with Ly$\alpha$; (2) \ion{Mg}{ii} with \ion{C}{iv}; (3) \ion{Mg}{ii} with \ion{C}{iii]} and (4) \ion{C}{iv} with Ly$\alpha$ (see section \ref{sec:app_results}).}
\label{fig:app_results}
\end{figure}
	
	Figure \ref{fig:app_results} illustrates the result of the cross-validation process for both the classical phase correlation and weighted phase correlation algorithms and further illustrates a comparison with the redshift predicted by the SDSS-III pipeline. We can readily see that the performances of the classical phase correlation algorithm are strongly degraded compared to the weighted version with a correlation factor of $0.557$ compared to $0.984$ and a ratio of observations having  $\left|\Delta z\right| < 0.05$ of $0.838$ compared to $0.992$, respectively. These differences mainly come from the previously mentioned problem of border flattening that translates into frequent emission line mismatches and into errors coming from the difficulty that the algorithm has in order to extrapolate the regions surrounding the Ly$\alpha$ and H$\alpha$ emission lines. This difficulty arise because of the prominence of these lines as well as because of the high correlation they have with the other emission lines \citep{yip2004}. As a consequence, the algorithm is often constrained to consider the Ly$\alpha$ or H$\alpha$ lines to be embedded within the observed spectra which graphically results in a gap around $0.4 < \bar{z} < 2.12$. Note that the systematic errors occurring at $\bar{z} \sim 0.4$ and at $\bar{z} \sim 2.12$ can be attributed to the fitting of these specific emissions lines to the residual spectrograph edge effects --particularly significant within the low SNR spectra-- and that these errors account for $\sim 2\%$ of the observations having $\left|\Delta z\right| \geq 0.05$.
	
\begin{figure}
\includegraphics{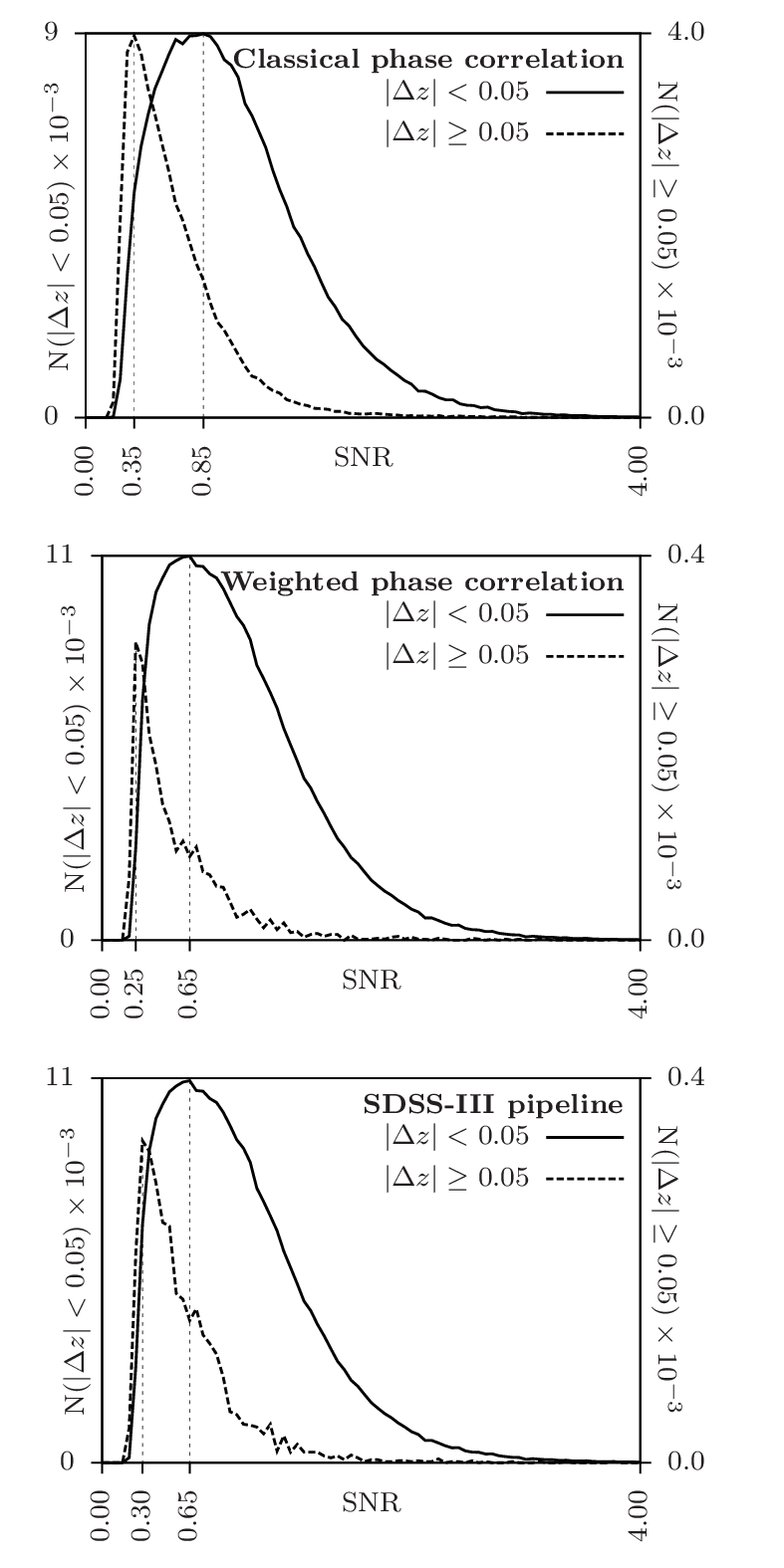}
\caption{Histograms of the SNR distribution amongst the observations having $\abs{\Delta z} < 0.05$ and those having $\abs{\Delta z} \geq 0.05$ for the cases of the classical phase correlation; of the weighted phase correlation and of the SDSS pipeline output. A bin width of $\Delta \mathrm{SNR} = 0.1$ was used in each of these graphs. Also highlighted is the position of the maximal peak of each histogram.}
\label{fig:app_snr_compare}
\end{figure}

	Investigation of the most significant errors coming from the emission lines mismatch, illustrated in figure \ref{fig:app_results}, shows that the latter can be modelled as a linear relation between the predicted redshift and the effective redshift. Indeed, if we consider an emission line observed at wavelength $\lambda$ and falsely considered to stand at a restframe wavelength $\lambda_\mathrm{f}$ instead of $\lambda_\mathrm{t}$, we will have that the predicted redshift, $z_\mathrm{f}$, can be related to the effective redshift, $z_\mathrm{t}$, through
\begin{equation}
\frac{z_\mathrm{f}+1}{z_\mathrm{t}+1} = \frac{\lambda_\mathrm{t}}{\lambda_\mathrm{f}}.
\label{eq:zt_vs_zf}
\end{equation}

	These mismatches do not constitute, in themselves, real cases of degeneracy regarding our $\chi_r^2$ and $Z_{\mathrm{score}}$ selection criteria. Indeed, each of the configuration mentioned within figure \ref{fig:app_results} has unconfused emission lines that make the resulting redshift unambiguous. Rather, the observed degeneracies also come from the low SNR of the observed spectra. Figure \ref{fig:app_snr_compare} illustrates the distribution of the SNR of both the observation having $\abs{\Delta z} \geq 0.05$ and those having $\abs{\Delta z} < 0.05$ for our three cases of study. We notice that for all three cases, the SNR of the maximal peak of the fair redshifts estimate is approximately twice the one of the erroneous ones, this is especially significant in the cases of the weighted phase correlation and of the SDSS-III pipeline where the errors come nearly exclusively from this line mismatch problem. Furthermore, a visual inspection of these degenerated spectra shows both potential redshifts to be undistinguishable from one another in most of the cases and thus constituting \textit{in fine} effective cases of degeneracy. Consequently, some of the low SNR spectra will unavoidably have ambiguous redshift estimates that will stand in well specific regions defined by equation \ref{eq:zt_vs_zf}. Nevertheless, these will be easily identified as having a low $Z_{\mathrm{score}}$ and/or a low redshift confidence (see section \ref{sec:discussion_zerror}).

	Finally, we may notice that our implementation seems to have a better tolerance to noise compared to the SDSS-III implementation (i.e. see within figure \ref{fig:app_snr_compare} the lower peak of the erroneous curve as well as its globally smaller width). This higher tolerance does not come from differences in the algorithms since both implementations are based on the sole solution to equation \ref{eq:wcorr_chi2}, but either: (1) from the higher number of PCA components we used (11 compared to 4); (2) from the fact that the components we used were more suitable in order to represent the observed spectra or (3) from the fact that the redshifts coming from the visual inspection procedure are also subject to errors, especially since we are concerned with low SNR spectra where degeneracy may occur.
	
	In order to reject the fact that this higher tolerance comes from the larger number of components we used, we repeat the described cross-validation procedure by using only three components (plus mean observation) instead of ten. The results of this configuration lead us to the same conclusions with a correlation factor of $0.976$ (compared to $0.967$) and a ratio of observations having  $\left|\Delta z\right| < 0.05$ of $0.989$ (compared to $0.988$). Although the differences in the erroneous SNR curves are less perceptible, it still remains globally sharper. Furthermore, we have to mention that within the SDSS-III pipeline, no more than four principal components were used because any larger number of components would make the error higher. In regard to this point and to the fact that we succeed in getting good predictions using 11 components, we might suppose, in anyway, that the components we used were of higher quality in order to model this specific dataset. Nevertheless, let us mention that we cannot totally reject the hypothesis according to which this better tolerance comes from a fortuitous statistical fluctuation itself produced by the degeneracy occurring during the visual inspection of some low SNR spectra.
	
\section{Discussion}
\label{sec:discussion}
\subsection{Redshift confidence \& uncertainty estimation}
\label{sec:discussion_zerror}

	In order for the derived redshift to be effectively used within subsequent scientific applications it is mandatory for it to come along with an estimation on its uncertainty and to have a confidence level that the chosen redshift is indeed in the vicinity of the real redshift. To make it clear, we may have a redshift estimation with a reasonable uncertainty (e.g. $z = 2.31\pm10^{-3}$) but being degenerated such that we are not sure that it stands in the neighbourhood of the effective redshift. Fortunately, the computed CCF offers us simple and efficient ways to evaluate both the redshift uncertainty as well as the confidence we can set on it.
	
	Generally speaking, we know that for a sufficiently large sample of observed points, the $\chi^2$ map defined in the parameters $\left\lbrace a_1,\cdots, a_n \right\rbrace$ can be approximated in the neighbourhood of the global minimum, $\left\lbrace a_1^\star,\cdots, a_n^\star \right\rbrace$, as
\begin{equation}
\chi^2(a_i) \approx \frac{\left(a_i-a_i^\star\right)^2}{\sigma^2(a_i^\star)} + C,
\label{eq:chi2_map_approx}
\end{equation}
where $C$ is a function depending on $a_j$, $j \neq i$ and thus considered here as a constant. In other words, the approximation of the $\chi^2$ map near a global minimum can be evaluated for each of the parameters independently from the others as a simple quadratic curve whose curvature depends on the uncertainty of the varying parameter. As a consequence, if $\chi^2(a_i)$ increases by one compared to the optimal $\chi^2$, then we will have that $\sigma^2(a_i) = \left(a_i-a_i^\star\right)^2$. The reader may find in \citet[section 8.1]{bevington2003} more informations about the variation of the $\chi^2$ near the optimum and more particularly about the derivation of equation \ref{eq:chi2_map_approx}. Regarding the uncertainty on the predicted redshifts, we used a second order polynomial such as to fit the optimal peak of the CCF, $Z$, and derived its associated uncertainty\footnote{Beware that the shift value corresponding to the uncertainty will have an associated decrease by one compared to the maximal peak of the CCF.}, $\sigma(Z)$. We then use the propagation of the uncertainty such as to get the error on the estimated redshift
\begin{equation}
\sigma(z) = (z+1) \sigma(Z) s \log b,
\end{equation}
where $b$ is the base of the logarithmic scale we used (in our case $b = 10$) and $s$ is the sampling of the spectra on this logarithmic scale (in our case $s = \Delta \log_{10} \lambda = 10^{-4}$).

	Secondly, we have to evaluate the confidence we can have on the predicted redshift. First estimators of this confidence are the already mentioned $Z_{\mathrm{score}}(z)$ and $\chi_r^2(z)$ (see section \ref{sec:app_proc_desc}). Indeed, a secure estimate will typically have $Z_{\mathrm{score}}(z) \approx \chi_r^2(z) \approx 1$. Unfortunately, these estimators do not take into account the potential ambiguity that might be present during the selection of the CCF peak associated with the predicted redshift. In order to tackle this lack, we defined the chi-squared difference associated with a redshift estimate, $z_i$, as
\begin{equation}
\Delta \chi_r^2(z_i) = \min \abs{\chi_r^2(z_i) - \chi_r^2(z_j)},\qquad \forall j \neq i,
\end{equation}
where each $z_j$ corresponds to a redshift associated with a peak selected within the CCF. We may notice that any redshift being unsure due to the ambiguity in the CCF peak selection will now be marked as having $\Delta \chi_r^2(z) \approx 0$. Also note that compared to \cite{bolton2012}, we use the distance between all $\chi_r^2$ and not only those for which $\chi_r^2(z_i) \geq \chi_r^2(z_j)$ because we adopt the hypothesis that any solution having $\chi_r^2(z_i) < \chi_r^2(z_j)$ might have been falsely rejected while being a valid solution.

\subsection{Dealing with zero weights}

	It commonly happens for the weight matrix, $\mat{W}$ from equation \ref{eq:wcorr_chi2}, to have a lot of successive weights set to zero, this is especially true if we consider that the observation can be padded such as either to match the size of the templates or to deal with the periodic nature of the phase correlation. Additionally, nothing prevents us from shifting the observation in a circular way such as to have this set of successive zeroes being in the first rows of the weights matrix and to later get rid of this artificial shift by sliding back the CCF. This is particularly interesting if we have a number of zeroed weights equal to --or greater than-- the number of components we used, $\ncomp$. In this case, the $\ncomp$ first rows of the matrix $\mat{\shifted{X}}$ and of the vector $\vec{y}$ used within the factorized QR algorithm with lookup tables will all be equal to zero and as a consequence, none of the $\mat{\shifted{X}_i^\prime}$'s, as well as none of the $\vec{y_i^\prime}$'s have to be computed. Differently stated, in addition to the building of the lookup tables and their updates, we solely have to compute $\mat{R}_{ij}$ and $b_i(Z)$ through equations \ref{eq:wcorr_qip_xip} and \ref{eq:wcorr_qip_yip}:
\begin{equation}
\mat{R}_{ij} = - \frac{\mat{\shifted{L}}_{ij}}{\sqrt{\mat{\shifted{L}}_{ii}}}
\label{eq:discussion_rij_zopt}
\end{equation}
and
\begin{equation}
b_i(Z) = - \dfrac{\vec{\shifted{l}}_{i}}{\sqrt{\mat{\shifted{L}}_{ii}}}.
\label{eq:discussion_biz_zopt}
\end{equation}
This allows us to greatly simplify our algorithm and leads to execution times of $0.082 \pm 0.001$s for the case $N = 10^4$, $\ncomp = 10$ and of $0.912\pm 0.026$s for the case $N = 10^5$, $\ncomp = 10$. A rough comparison shows these execution times to be twice faster than those presented at the end of section \ref{sec:wcorr_orth_factorized_lookup}.

\subsection{Templates weighting}

	Although, the weighting of the observed spectra is the most important regarding the redshift determination of QSOs, one might also want to have a template weighting such as, for example, to highlight some patterns or to reflect the fact that these templates often come along with their own uncertainties. To this aim, we plug into equation \ref{eq:wcorr_chi2}, the diagonal matrix of weights associated with the template observations, $\mat{W_T}$, that is
\begin{equation}
\chi^2(Z) = \left\|\mat{\shifted{W}_T}\mat{W}\vec{s}-\mat{W}\mat{\shifted{W}_T}\mat{\shifted{T}}\vec{a}(Z)\right\|^2 = \left\|\vec{\shifted{y}}-\mat{\shifted{X}}\vec{a}(Z)\right\|^2.
\end{equation}
After orthogonalization of the matrix $\mat{\shifted{X}} = \mat{\shifted{Q}} \mat{\shifted{R}}$, we get to
\begin{equation}
\chi^2(Z) = \left\|\vec{\shifted{y}}\right\|^2- \left\|\vec{b}(Z)\right\|^2,
\label{eq:wtemplate_chi2_orth}
\end{equation}
with the first $\ncomp$ elements of $\vec{b}(Z)$ being equal to the first $\ncomp$ elements of $\matt{\shifted{Q}}\vec{\shifted{y}}$. We can already note that since $\vec{\shifted{y}}$ is now shift-dependent, the knowledge of $\vec{b}(Z)$ alone is no more sufficient in order to find the optimal shift such that $\chi^2(Z)$ must be explicitly evaluated through equation \ref{eq:wtemplate_chi2_orth}.

	Computation of the first $\ncomp$ elements of $\vec{b}(Z)$ is straightforwardly done using the procedure described in section \ref{sec:wcorr_orth_factorized_lookup} with $\vec{\shifted{y}}$ replacing $\vec{y}$ and both lookup tables given by
\begin{equation}
\mat{\shifted{L}}_{ij} = \ifourp{\conj{\fourp{\mat{W_T}^2\left(\matc{T}{i}\eprod\matc{T}{j}\right)}} \eprod \fourp{\mat{W}^2}}_Z
\label{eq:wtemplate_lookup}
\end{equation}
and
\begin{equation}
\vec{\shifted{l}}_i = \ifourp{\conj{\fourp{\mat{W_T}^2\matc{T}{i}}} \eprod \fourp{\mat{W}^2\vec{s}}}_Z.
\label{eq:wtemplate_ylookup}
\end{equation}
Finally, we will have that each $\left\|\vec{\shifted{y}}\right\|^2$ will be given by
\begin{equation}
\left\|\vec{\shifted{y}}\right\|^2 = \ifourp{\fourp{\mat{W_T}^2}\eprod\fourp{\mat{W}^2\vec{s}^2}}_Z.
\end{equation}

\section{Conclusions}
\label{sec:conclusion}

We have presented a new method for computing the weighted phase correlation of an observed input signal against not necessarily orthogonal templates. This method is found to be the preferred alternative to the classical phase correlation in the case of input observations having a limited coverage and/or having very distinct weights. The implementation of this method is based on a weighted chi-squared problem solved through a highly modified version of the QR orthogonalization algorithm designed to take benefit of the performances of the fast Fourier transform such as to compute the numerous inner products present within the original QR algorithm. This implementation provides us with a numerically stable algorithm having a linearithmic time complexity that makes it affordable for the tight spectral processing of QSOs within the Gaia mission.

	We have presented a complete application of this method to the case of the redshift determination of type I/II QSOs coming from the SDSS DR12 quasar catalog through a two-fold cross-validation procedure. This application is based on templates coming from the weighted principal components analysis decomposition of independent spectra coming from the same catalog. We described in detail the reduction of those input spectra as well as the method we used in order to select the most probable redshift amongst the set of possible ones. Results of this cross-validation show our method to be the one of predilection for QSO redshift determination and is comparable to the SDSS-III pipeline output while not being a $\bigo{N^2}$ process.

	Finally, we showed how we can get both the uncertainty on the predicted redshift as well as the confidence we can set on it. We further  discuss two extensions of our method, namely: the time saving we can get if having a sufficient number of successive zeroed weights and the weighting of the template observations.
	
	A free implementation of the described algorithm has been released under the GNU Public License\footnote{\href{http://www.gnu.org/licenses/gpl-3.0.txt}{http://www.gnu.org/licenses/gpl-3.0.txt}} and can be freely downloaded at \href{https://github.com/ldelchambre/wcorrQRL}{https://github.com/ldelchambre/wcorrQRL}.

\section*{Acknowledgements}
The author acknowledges support from the ESA PRODEX Programme `Gaia-DPAC QSOs' and from the Belgian Federal Science Policy Office.

Funding for SDSS-III has been provided by the Alfred P. Sloan Foundation, the Participating Institutions, the National Science Foundation and the U.S. Department of Energy Office of Science. The SDSS-III web site is http://www.sdss3.org/.

SDSS-III is managed by the Astrophysical Research Consortium for the Participating Institutions of the SDSS-III Collaboration including the University of Arizona, the Brazilian Participation Group, Brookhaven National Laboratory, Carnegie Mellon University, University of Florida, the French Participation Group, the German Participation Group, Harvard University, the Instituto de Astrofisica de Canarias, the Michigan State/Notre Dame/JINA Participation Group, Johns Hopkins University, Lawrence Berkeley National Laboratory, Max Planck Institute for Astrophysics, Max Planck Institute for Extraterrestrial Physics, New Mexico State University, New York University, Ohio State University, Pennsylvania State University, University of Portsmouth, Princeton University, the Spanish Participation Group, University of Tokyo, University of Utah, Vanderbilt University, University of Virginia, University of Washington and Yale University.

\bsp	
\label{lastpage}
\end{document}